\documentclass[a4paper,final,twocolumn,floatfix]{revtex4-1}

\pdfoutput=1

\usepackage{amsmath,amssymb}
\usepackage{graphicx}

\newcommand{\reffig}[1]{FIG.~\ref{#1}}

\newcommand{\refsec}[1]{Section~\ref{#1}}

\begin{document}

\title{Slow and velocity-tunable beams of metastable He$_2$ by multistage Zeeman deceleration}
\author{Michael Motsch}
\author{Paul Jansen}
\author{Josef A. Agner}
\author{Hansj\"urg Schmutz}
\author{Fr\'ed\'eric Merkt}
\affiliation{Laboratorium f\"ur Physikalische Chemie, ETH Z\"urich, CH-8093, Switzerland}

\begin{abstract}

$\mathrm{He_2}$ molecules in the metastable $\mathrm{a\,^3\Sigma_\mathrm{u}^{+}}$ state have been generated by striking a discharge in a supersonic expansion of helium gas from a pulsed valve. When operating the pulsed valve at room temperature, 77\,K, and 10\,K, the mean velocity of the supersonic beam was measured to be 1900\,m/s, 980\,m/s, and 530\,m/s, with longitudinal velocity distributions corresponding to temperatures of 4\,K, 1.9\,K, and 1.8\,K, respectively. The characterization of the population distribution among the different rotational levels of the $\mathrm{a\,^3\Sigma_\mathrm{u}^{+}}$ state by high-resolution photoelectron and photoionization spectroscopy indicated a rotational temperature of about 150\,K for the beam formed by expansion from the room-temperature valve and a bimodal distribution for the beam produced with the valve held at 10\,K, with rotational levels up to $N^{\prime\prime}=21$ being populated. A 55-stage Zeeman decelerator operated in a phase-stable manner in the longitudinal and transverse dimensions was used to further reduce the beam velocity and tune it in the range between 100 and 150\, m/s. The internal-state distribution of the decelerated sample was determined by recording and analyzing the photoionization spectrum in the region of the lowest ionization threshold, where it is dominated by resonances corresponding to autoionizing $n$p Rydberg states belonging to series converging to the different rotational levels of the $\mathrm{X\,^2\Sigma_\mathrm{u}^{+}}$ ground state of He$_2^+$. The deceleration process did not reveal any rotational state selectivity, but eliminated molecules in spin-rotational sublevels with $J^{\prime\prime}=N^{\prime\prime}$ from the beam, $J^{\prime\prime}$ and $N^{\prime\prime}$ being the total and the rotational angular momentum quantum number, respectively. The lack of rotational state selectivity is attributed to the fact that the Paschen-Back regime of the Zeeman effect in the rotational levels of the $\mathrm{a\,^3\Sigma_\mathrm{u}^{+}}$ state of $\mathrm{He_2}$ is already reached at fields of only 0.1\,T.
\end{abstract}

\maketitle

%%%%%%%%%%%%%%%%%%%%%%%%%%%%%%%%%%%%%%%%%%%%%%%%%%%%%%%%%%%%%%%%%%%%%%%%%%%
% INTRODUCTION
%%%%%%%%%%%%%%%%%%%%%%%%%%%%%%%%%%%%%%%%%%%%%%%%%%%%%%%%%%%%%%%%%%%%%%%%%%%

\section{Introduction}
\label{sec:intro}

In the last decade, several methods have become available to prepare molecules in the gas phase at low temperatures~\cite{bell09b, hogan11a, vandemeerakker12a, narevicius12a}. Going beyond thermalization with a cold environment, these methods allow for a precise control of the external and internal degrees of freedom in molecular samples. Cold molecules with translational temperatures in the range 100\,mK--1\,K may then find applications in precision spectroscopy~\cite{vanveldhoven02a, hudson11a} and in the study of chemical reactivity at very low collision energies and very high energy resolution~\cite{gilijamse06a, kirste12a}. Alternatively, they may also represent the starting point for further cooling steps toward the ultracold regime below 1\,mK~\cite{shuman09a, shuman10a, zeppenfeld12a, barry12a, zhelyazkova13aarxiv}.
The choice of the cooling method depends on the properties of the species of interest~\cite{vandemeerakker08a, hogan11a, lemeshko13a}. In the case of paramagnetic atoms and molecules which can be entrained in a supersonic beam, multistage Zeeman deceleration~\cite{vanhaecke07a} and related techniques~\cite{trimeche11a, lavertofir11a} are the methods of choice: The phase-space density of atoms and molecules in supersonic beams is high and is preserved by a phase-stable operation of the decelerator~\cite{bethlem02a, wiederkehr10b}, the method can be quantum-state selective~\cite{wiederkehr12a}, the final velocity of the particles can be controlled and tuned over a wide range~\cite{narevicius08a, wiederkehr11a, momose13a}, and the decelerated particles can be loaded into a magnetic trap~\cite{hogan08d, wiederkehr10a} for observation of slow decay processes or spectroscopic measurements requiring long observation times.

We describe here the use of multistage Zeeman deceleration to generate slow and velocity-tunable beams of He$_2$ molecules in the metastable $\mathrm{a\,^3\Sigma_u^+}$  state (referred to as He$_2^*$ hereafter) in a setup tailored for future applications in precision spectroscopy. We also demonstrate the possibility to use the cold molecular samples to study the rovibrational energy-level structure of the $\mathrm{a\,^3\Sigma_u^+}$ state of He$_2$ and the $\mathrm{X\,^2\Sigma_u^+}$ ground state of He$_2^+$ by photoionization and photoelectron spectroscopic methods. Being three- and four-electron systems, respectively, He$_2$ and He$_2^+$ represent attractive systems to test the accuracy of \textit{ab-initio} quantum-chemical methods. In the two-electron systems H$_2$, HD, and D$_2$, the results of \textit{ab-initio} calculations of dissociation energies and rovibrational energy levels which include adiabatic and nonadiabatic corrections to the Born-Oppenheimer approximation as well as quantum-electrodynamics corrections up to the leading (one-loop) contribution to the term proportional to $\alpha^4$ were found to be in agreement with experimental results within the combined uncertainties of the experimental and theoretical determinations~\cite{liu09b, liu10a, sprecher10a, pachucki09a, piszczatowski09a, dickenson12a, sprecher11a, dickenson13a}. Such calculations become increasingly challenging as the number of electrons increases, and the magnitudes of the relativistic and quantum-electrodynamics corrections are not accurately known for He$_2^+$. In a recent high-resolution spectroscopic study of the threshold ionization spectrum of He$_2^*$, the level spacings between the lowest four rotational levels of He$_2^+$ could be determined at an accuracy of 100\,MHz~\cite{sprecher14a}. Comparison with the latest \textit{ab-initio} calculations neglecting quantum-electrodynamics corrections~\cite{tung12a} did not reveal significant deviations from the experimental results at this level of accuracy, except perhaps for the highest rotational level observed experimentally. In future, we would like to improve the accuracy of these measurements by using a slow beam of metastable He$_2$, which motivated us to develop a He$_2^*$ source better suited to precision spectroscopic experiments.

As a starting point, a supersonic beam of $\mathrm{He}_2^*$ was produced with a pulsed source operated at cryogenic temperatures. The velocity of the molecules was subsequently reduced to $\approx$100\,m/s by coupling the cryogenic beam source to a multistage Zeeman decelerator. From pulsed-field-ionization zero-kinetic-energy (PFI-ZEKE) photoelectron spectra of the $\mathrm{X^+\,^2\Sigma_u^+ \leftarrow a\,^3\Sigma_u^+}$ transition, we have determined the internal-state distribution of the  cold molecules in the beam and found rotational states up to $N^{\prime\prime}=21$ to be populated. Deceleration experiments with these molecules enabled the study of the rotational state selectivity of the deceleration process and a comparison with the deceleration of molecular oxygen ($\mathrm{O}_2$) in the  $\mathrm{X\,^3\Sigma_g^-}$ state, for which full selectivity of the spin-rotational state $\left|N=1,J=2,M_J=2\right>$ was achieved~\cite{wiederkehr12a}.

%%%%%%%%%%%%%%%%%%%%%%%%%%%%%%%%%%%%%%%%%%%%%%%%%%%%%%%%%%%%%%%%%%%%%%%%%%%
% EXPERIMENTAL SETUP
%%%%%%%%%%%%%%%%%%%%%%%%%%%%%%%%%%%%%%%%%%%%%%%%%%%%%%%%%%%%%%%%%%%%%%%%%%%

\section{Experiment}
\label{sec:setup}

$\mathrm{He}_2^*$ is formed in a gas of pure helium under conditions where the density allows for three-body collisions on the relevant time scales and collision partners of sufficiently high energy are available to excite helium atoms to their metastable states. Such conditions typically prevail in plasmas, such as those generated by striking an electric discharge through helium gas~\cite{chabalowski89a, raunhardt08a}. The large ratio of magnetic moment to mass makes $\mathrm{He}_2^*$ particularly attractive for multistage Zeeman deceleration. Moreover, the predicted radiative lifetime of 18\,s of $\mathrm{He}_2^*$~\cite{chabalowski89a} is sufficiently long that no decay takes place on the millisecond timescale of the deceleration process. However, the high velocity (about 2000\,m/s) of a supersonic expansion of pure He from a reservoir held at room temperature poses a problem. For the multistage Zeeman decelerator used in our experiments, the deceleration-solenoid geometry and the 8\,$\mu$s switching times of the pulsed deceleration magnetic fields limit the maximal initial velocity of the particles to be decelerated to about 700\,m/s~\cite{wiederkehr10b, hogan11a}. The common approach to produce low-velocity supersonic beams consisting of  seeding the species of interest in a heavy rare gas is not an option in the case of $\mathrm{He}^*$ and $\mathrm{He}_2^*$ because these metastable species carry sufficient internal energy to cause the Penning ionization of all other rare gases~\cite{siska93a}. The operation of the pulsed-valve assembly, including discharge electrodes, at cryogenic temperatures is therefore the only viable option to reduce the initial velocity of the supersonic beam to below 700\,m/s.

\begin{figure}
\centering
\includegraphics{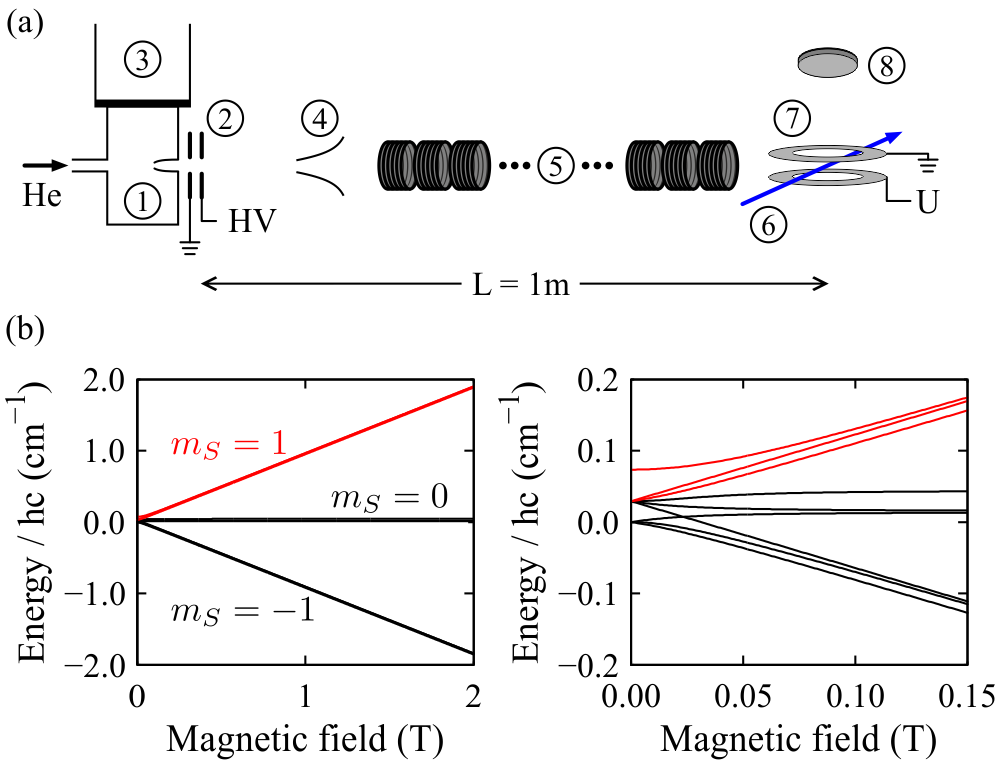}
\caption{(a) Schematic representation of the experimental setup (not to scale). 1: Pulsed valve body and orifice, 2: Discharge electrodes, 3: Cryogenic-liquid container, 4: skimmer, 5: multistage Zeeman decelerator, 6: UV laser beam, 7: extraction electrodes, 8: MCP. The supersonic beam propagation direction, the UV laser beam, and the ion extraction direction are mutually orthogonal.
(b) Zeeman effect in the $N^{\prime\prime}=1$ rotational ground state of $\mathrm{He}_2$ $\mathrm{(a\,^3\Sigma_u^+)}$. The low-field-seeking states, for which the deceleration pulse sequence is optimized, are indicated in red.}
\label{fig:setup}
\end{figure}

To produce slow and velocity-tunable beams of metastable helium molecules at low speed, we have combined a pulsed cryogenic supersonic beam source with a multistage Zeeman decelerator in the experimental setup depicted schematically in \reffig{fig:setup}~(a). The setup consists of the pulsed-valve assembly with discharge electrodes, the actual multistage Zeeman decelerator, and a detection region where the metastable molecules are photoionized with UV laser radiation and the ions or electrons are extracted in the direction perpendicular to the beam propagation axis toward a microchannel-plate (MCP) detector. The Zeeman decelerator was described in detail in Ref.~\cite{wiederkehr11a}. Consequently, only the beam source is presented in detail in the following description of the experimental setup and procedure.

\subsection{The cryogenic discharge source}
\label{sec:setup:cryo}

The pulsed supersonic beam is created by expanding helium gas into vacuum from a reservoir held at high stagnation pressure (typically 2--6\,bar) through the 250\,$\mu$m-diameter ceramic orifice of a modified Even-Lavie valve~\cite{even00a}. To efficiently cool the valve, its entire housing, originally made of stainless steel, was replaced by copper components. The valve assembly is mounted onto the copper baseplate of a $\approx$0.7\,l cylindrical container which can be filled with cryogenic liquids, either liquid nitrogen for operation at 77\,K, or liquid helium for operation down to 4\,K. The temperature of the valve assembly is monitored with a silicon diode (Lakeshore DT-670) directly attached to it. Throughout this paper, we refer to the temperature measured with this sensor as the ``source temperature''.

Starting with the valve assembly at room temperature, it takes about two hours until the valve and discharge operate stably at about 10\,K. Once the desired final temperature is reached, the nozzle-assembly temperature is actively stabilized with a resistive heater attached to the valve. With feedback from a closed-loop proportional-integral-derivative (PID) controller (Lakeshore 336), the source temperature fluctuates by less than 0.1\,K and conditions remain stable for several hours.

The metastable helium molecules are generated by striking a discharge and forming a plasma in the expansion region. For this purpose, the expanding gas passes through a 10\,mm-long, 1.6\,mm-diameter channel defined by two cylindrical electrodes directly attached to the front plate of the valve and separated by an electrically insulating sapphire spacer (2\,mm length, inner diameter of cylindrical channel 2\,mm). The discharge is ignited by applying a typically 5\,$\mu$s long, $-600$\,V pulse to the front electrode as soon as the density of the helium beam in the channel is sufficiently high. In the discharge, the metastable $\mathrm{(1s)(2s)}$ $\mathrm{^1S_0}$ and $\mathrm{^3S_1}$ states of atomic helium are populated, and the metastable helium dimer molecules are formed. To facilitate ignition of the discharge during the short duration of the gas pulse (typically a few tens of microseconds), the discharge is seeded with electrons emitted from a hot tungsten filament~\cite{halfmann00a, wiederkehr11a} and attracted toward the channel by a $+200$\,V bias on the front electrode. Without the filament, ignition of the discharge required much higher electric tensions, which unnecessarily heated the beam. The discharge and nozzle operation parameters (electric tension, timing, stagnation pressure) were optimized by monitoring the yield of $\mathrm{He}_2^*$ with the methods described in \refsec{sec:setup:detection}.

We have also tested other discharge-electrode designs (e.g., high electric tension applied to a sharp tip in the expansion region~\cite{halfmann00a, wiederkehr11a}, cylindrical electrodes surrounding a glass capillary) and found the best yield of $\mathrm{He}_2^*$ with the channel described above. All discharge configurations produced beams of metastable helium atoms, but $\mathrm{He_2^*}$ was only observed in the configuration described above.

In the experiment, we observed a strong influence of the discharge electrode temperature on the final velocity of the beam of metastable helium atoms and molecules. This observation may be explained by the discharge electrodes being part of the channel through which the gas expands, thus constituting a virtual gas reservoir for the expansion and beam-formation process. To cool this critical part of the setup to the lowest possible temperatures, the discharge assembly was mechanically connected and thermally linked to the copper front plate of the valve body. Moreover, the valve and discharge assembly were surrounded by a thermal shield cooled with liquid nitrogen, which served the purpose of minimizing the heat load of thermal radiation from the hot filament and from components of the vacuum chamber at room temperature.

For fast data acquisition in spectroscopic experiments, the source can be operated at the full repetition rate of 25\,Hz of the Nd:YAG-pumped dye-laser system used to detect the metastable molecules. In this case, a 650\,l/s (nominal pumping speed for He) turbo molecular pump sufficed to maintain the pressure in the source vacuum chamber in the $10^{-5}$\,mbar  range. After the formation of He$_2^*$ molecules, the supersonic beam passed through the 2\,mm-diameter hole of a skimmer, which separates the source chamber from the multistage Zeeman decelerator or the photoelectron spectrometer used to characterize the beam properties and guarantees efficient differential pumping.

\subsection{The multistage Zeeman decelerator}
\label{sec:setup:zeeman}

The multistage Zeeman decelerator used in the present work consists of an array of deceleration solenoids to which current pulses are applied and was described in Ref.~\cite{wiederkehr11a}. The deceleration results from the force
\begin{equation}
\vec{f}_\mathrm{Z}=-\nabla E_\mathrm{Z} = \nabla(\vec{\mu}_\mathrm{mag}\cdot \vec{B})
\label{force}
\end{equation}
that acts on a paramagnetic atom or molecule with magnetic moment $\vec{\mu}_\mathrm{mag}$ and Zeeman energy shift $E_\mathrm{Z}$ in the presence of an inhomogeneous magnetic field $\vec{B}$. Particle-trajectory simulations indicated that 55 deceleration stages operated at a current of 300\,A suffice to fully decelerate a beam of $\mathrm{He}_2^*$ with an initial velocity of about 500\,m/s. The 55 solenoids (length 7.2\,mm, 64 windings in four layers, 7.0\,mm inner diameter) used as deceleration stages were grouped in two modules of each twelve stages and one module of 31 stages. These modules were separated by pumping towers, as depicted in Fig.~2 of Ref.~\cite{wiederkehr11a}.

The Zeeman effect in the lowest rotational level $(N^{\prime\prime}=1)$ of the metastable $\mathrm{a\,^3\Sigma_u^+}$ state of He$_2$ is depicted in \reffig{fig:setup}~(b), where the low-field-seeking magnetic sublevels ($m_S=1$, drawn in red) suitable for deceleration experiments have a magnetic moment corresponding to two Bohr magnetons. Because the spin-rotation coupling constant of low vibrational levels of the $\mathrm{a\,^3\Sigma_u^+}$ state of He$_2$ is very small~\cite{focsa98a}, the Zeeman effect in higher rotational levels follows the same pattern as in the $N^{\prime\prime}=1$ level (see also discussion in \refsec{sec:conclusion}).
To generate beams of He$_2^*$ with velocities in the range 100--150\,m/s, the decelerator was operated at a repetition rate of 8.3\,Hz by applying 250\,A current pulses to the deceleration solenoids, resulting in maximal magnetic fields on the decelerator axis of about 1.8\,T and Zeeman shifts of about 1.6\,cm$^{-1}$ for the low-field-seeking magnetic sublevels. The pulse sequences were precalculated to reach the desired final velocities for molecules in these sublevels. Optimal conditions were obtained for phase angles between 35$^\circ$ and 45$^\circ$, as in our previous deceleration experiments with deuterium atoms~\cite{wiederkehr10b}, metastable Ne~\cite{wiederkehr11a}, and O$_2$~\cite{wiederkehr12a}.

\subsection{State-selective detection of the metastable He$_2$ molecules}
\label{sec:setup:detection}

\begin{figure}
\centering
\includegraphics{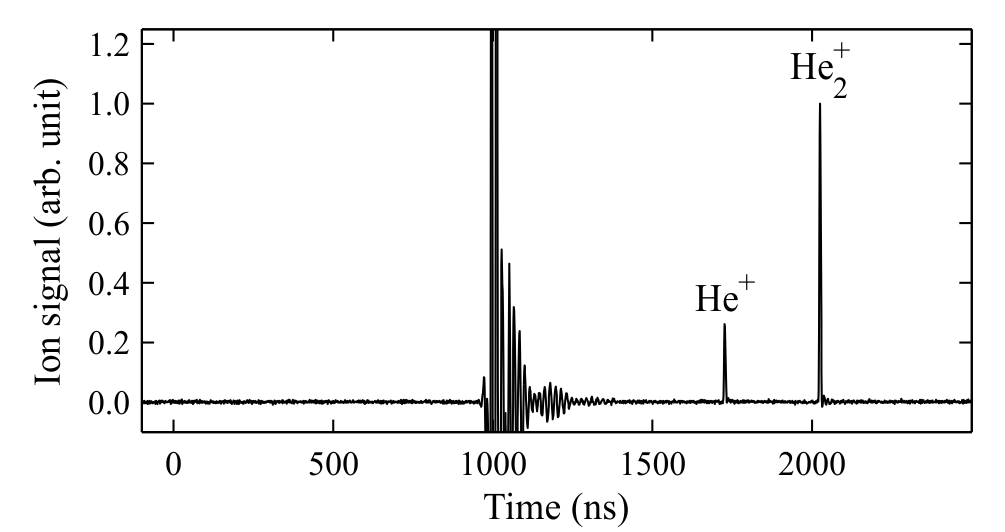}
\caption{Ion-TOF spectrum obtained following UV-laser photoionization at a wave number of 34467\,$\mathrm{cm^{-1}}$ and extracting the $\mathrm{He^+}$ and $\mathrm{He_2^+}$ ions with a pulsed extraction field applied  1\,$\mu$s after the laser pulse. The electronic noise between 1.0--1.25\,$\mu$s is an artifact originating from the electric-field switch-on process.}
\label{fig:iontof}
\end{figure}

The electric discharge described in \refsec{sec:setup:cryo} leads to the production of metastable He atoms in the $\mathrm{(1s)(2s)}$ $\mathrm{^1S_0}$ and $\mathrm{^3S_1}$ states and of metastable $\mathrm{He_2}$ molecules in the $\mathrm{a\,^3\Sigma_u^+}$ state. Because both atoms and molecules are entrained at the same velocity in the supersonic expansion, they cannot be distinguished on the basis of their  times of flight to a detector placed along the propagation axis of the supersonic beam.
To observe the atoms and the molecules separately, they were photoionized with UV laser radiation produced by second-harmonic generation of a Nd:YAG-pumped pulsed dye laser and the photoions were extracted through a field-free time-of-flight tube toward a microchannel-plate detector using an electric-field pulse in the range 1000--2000\,V/cm. The UV wave number (34467\,$\mathrm{cm^{-1}}$) was chosen to be above the ionization threshold of the metastable $\mathrm{a\,^3\Sigma_u^+}$ state of He$_2$, in a region free of autoionization resonances, and also above that of the metastable (1s)(2s) $^1\mathrm{S}_0$ state of He. Consequently, both He$^+$ and He$_2^+$ were observed in the ion time-of-flight (TOF) spectrum, as is illustrated in \reffig{fig:iontof}. The enhanced electronic noise visible in the time-of-flight spectrum between 1.0--1.25\,$\mu$s corresponds to the time at which the extraction field was applied, i.e., 1\,$\mu$s after the laser pulse.

To measure the photoionization spectra of He$_2^*$, the signal was integrated over a time window centered at the arrival time of $\mathrm{He_2^+}$ and recorded as a function of the excitation wave number, which was calibrated by recording the optogalvanic spectrum of Ne and the transmission through an etalon simultaneously with each spectrum. High-resolution photoelectron spectra of $\mathrm{He_2^*}$ were recorded using the technique of PFI-ZEKE photoelectron spectroscopy~\cite{reiser88a} following the same procedure as in our earlier study of the He$_2^+\; \mathrm{X^+\,^2\Sigma_u^+}\leftarrow \mathrm{He}_2^*\; \mathrm{a\,^3\Sigma_u^+}$ photoionizing transition~\cite{raunhardt08a}. The internal state distribution of the metastable molecules was deduced from the rotational structure of the photoelectron spectra and the analysis of the Rydberg series observed in the photoionization spectra, as explained in \refsec{sec:results:source}.

The flight times of the metastable molecules from the point where they were generated in the discharge to the point where they were photoionized by the UV laser (called ``neutral-TOF spectra'' hereafter) were obtained by monitoring the resulting $\mathrm{He_2^+}$ signal as a function of the delay time between the ignition of the discharge and the photoionization laser pulse. From the measured TOF distributions, the velocity distribution of the metastable molecule in the beam can be determined.

%%%%%%%%%%%%%%%%%%%%%%%%%%%%%%%%%%%%%%%%%%%%%%%%%%%%%%%%%%%%%%%%%%%%%%%%%%%
% RESULTS
%%%%%%%%%%%%%%%%%%%%%%%%%%%%%%%%%%%%%%%%%%%%%%%%%%%%%%%%%%%%%%%%%%%%%%%%%%%

\section{Results}
\label{sec:results}

%%%%%%%%%%%%%%%%%%%%%%%%%%%%%%%%%%%%%%%%%%%%%%%%%%%%%%%%%%%%%%%%%%%%%%%%%%%
% CHARACTERIZATION OF THE BEAM SOURCE
%%%%%%%%%%%%%%%%%%%%%%%%%%%%%%%%%%%%%%%%%%%%%%%%%%%%%%%%%%%%%%%%%%%%%%%%%%%

\subsection{Characterization of the supersonic beam}
\label{sec:results:source}

\begin{figure}
\centering
\includegraphics{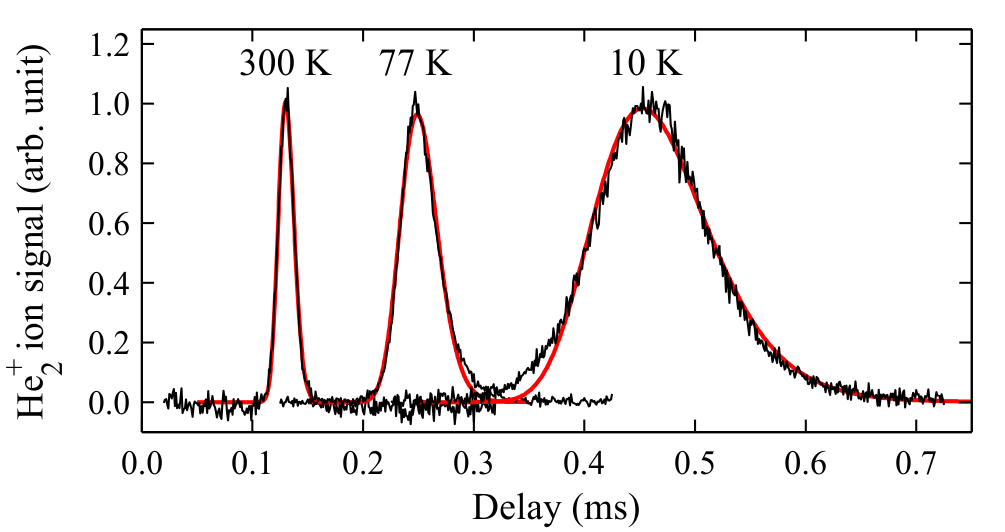}
\caption{He$_2^*$ neutral-TOF spectra obtained for temperatures of the nozzle assembly of 300\,K, 77\,K, and 10\,K. The red curves correspond to fits to the experimental data that led to the parameters summarized in Table~\ref{tab:velocity}. See text for details.}
\label{fig:metastabletof}
\end{figure}

The $\mathrm{He_2^*}$ neutral-TOF spectra obtained for nozzle-assembly temperatures of 300\,K, 77\,K and 10\,K by varying the delay between the discharge ignition pulse and the firing of the ionization laser are compared in \reffig{fig:metastabletof}. Analyzing these neutral-TOF distributions following the procedure described in Ref.~\cite{hogan11a, wiederkehr11a} enables one to extract, in each case, the average velocity of the molecular beam and the corresponding standard deviation, from which the translational temperature of the  He$_2^*$ sample can be estimated. From fits to the experimental time-of-flight profiles, the mean velocity and velocity spread of the beam are determined to be $\bar{v}$=(1875, 975, 530)\,m/s and $\sigma_{v}$=(95, 65, 60)\,m/s for source temperatures of (300, 77, 10)\,K. When the valve body is stabilized to 10\,K, the beam properties are sensitive to the discharge conditions, which results in day-to-day variations of the mean velocity by $\pm10$\,m/s. The results of this analysis are summarized in Table~\ref{tab:velocity} and indicate that the main effect of cooling the nozzle is a strong reduction of the average velocity of the He$_2^*$ molecules, the standard deviation being almost independent of the temperature of the nozzle assembly, particularly below 100\,K.

\begin{table}
\caption{Parameters describing the velocity distribution of the beam of $\mathrm{He_2^*}$ obtained from an analysis of the neutral-TOF spectra shown in \reffig{fig:metastabletof}.}
\begin{ruledtabular}
\begin{tabular}{lrrr}
$T_\mathrm{{Source}}$ (K) & $\bar{v}$ (m/s) & $\sigma_v$ (m/s) & $T$ (K) \\
300 & 1875 & 95 & 4.4\\
77 & 975 & 65 &  1.9\\
10 & 530 & 60 & 1.8\\
\end{tabular}
\end{ruledtabular}
\label{tab:velocity}
\end{table}

\begin{figure*}
\centering
\includegraphics{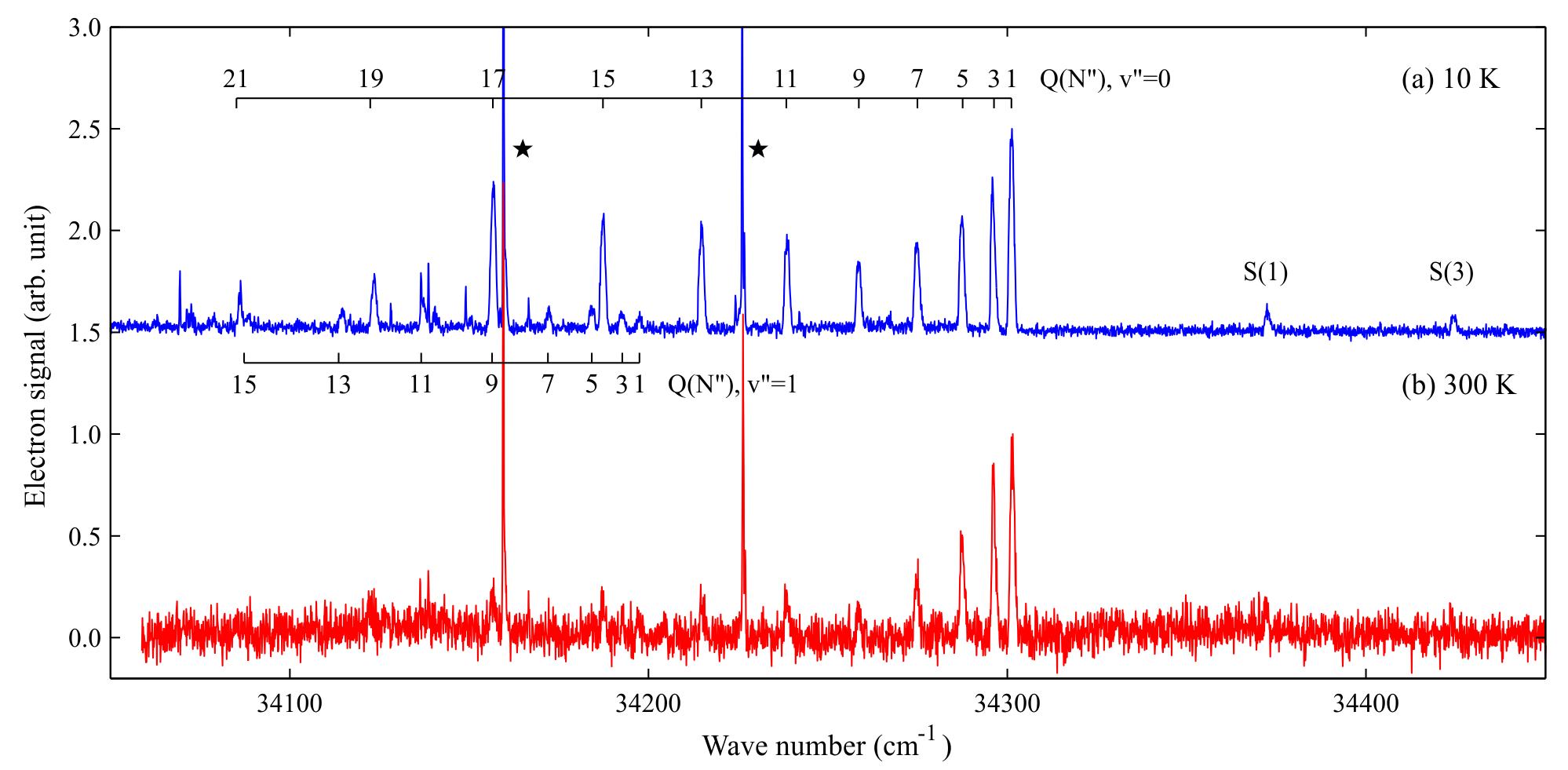}
\caption{PFI-ZEKE photoelectron spectra of the origin band of the $\mathrm{He_2^+\; X\,^2\Sigma_u^+ \leftarrow He_2^*\; a\,^3\Sigma_u^+}$  photoionizing transition obtained for nozzle-assembly temperatures of (a) 10\,K (blue trace), and (b) 300\,K (red trace). The numbers along the upper and lower assignment bars represent the values of the rotational quantum number $N^{\prime\prime}$ of $\mathrm{He_2^*}$ of the dominant Q-type-branch (i.e., $\Delta N=N^+-N^{\prime\prime}=0$) lines of the $v^+=0 \leftarrow v^{\prime\prime}=0$ and $v^+=1 \leftarrow v^{\prime\prime}=1$ bands, respectively. The weak lines designated as S(1) and S(3) are $\Delta N=N^+-N^{\prime\prime}=2$ transitions originating from the rotational levels $N^{\prime\prime}=1$ and 3, respectively. The two strong, sharp lines marked by asterisks correspond to autoionization resonances which strongly enhance the intensities of the O(3) and O(5) lines. The intensities in both spectra were normalized to the intensity of the Q(1) transition. See text for details.}
\label{fig:zeke}
\end{figure*}

The internal-state distribution of the molecular sample in the supersonic beam was determined from the rotational structure of the photoelectron spectrum of the origin band of the $\mathrm{He_2^+ \; X\,^2\Sigma_u^+ \leftarrow He_2^*\; a\,^3\Sigma_u^+}$  photoionizing transition. The photoelectron spectra obtained for nozzle-assembly temperatures of 10\,K and 300\,K are displayed in \reffig{fig:zeke}~(a) and (b), respectively. They were recorded using the technique of PFI-ZEKE photoelectron spectroscopy~\cite{reiser88a} by monitoring the delayed pulsed-field ionization of high Rydberg states (principal quantum number $n \ge 100$) located below the successive ionization thresholds of $\mathrm{He_2^*}$ as a function of the UV laser wave number. As explained in Ref.~\cite{raunhardt08a},  removal of the $3s\sigma$ electron leads to a dominant Q-type rotational branch in the PFI-ZEKE photoelectron spectrum of $\mathrm{He_2^*}$, corresponding to transitions between rotational levels of the neutral molecule and the ion having the same rotational quantum number, i.e., $\Delta N=N^+-N^{\prime\prime}=0$ ($N^{\prime\prime}$ and $N^+$ are the rotational quantum numbers of $\mathrm{He_2^*}$ and $\mathrm{He_2^+}$, respectively). The positions of the Q-type transitions of the $v^+=0 \leftarrow v^{\prime\prime}=0$ band for $N^{\prime\prime}$ values between 1 and 21 are indicated along the upper assignment bar at the top of \reffig{fig:zeke}~(a). Weak S-type lines (i.e., lines corresponding to $\Delta N=N^+-N^{\prime\prime}=2$) are also observable in \reffig{fig:zeke}, and so are two strong and sharp autoionization resonances which strongly enhance the intensity of the O(3)- and O(5)-branch lines as a result of rotational channel interactions, as discussed in detail in Ref.~\cite{raunhardt08a}. The very weak lines corresponding to the lower assignment bar form the Q-type branch of the $v^+=1 \leftarrow v^{\prime\prime}=1$ band. The observation of these lines indicates that the $v^{\prime\prime}=1$ level of the a $^3\Sigma_u^+$ state of He$_2$ is weakly populated in the supersonic beam.

The higher signal-to-noise ratio obtained at a nozzle-assembly temperature of 10\,K indicates that the density of He$_2^*$ is almost an order of magnitude larger than when the nozzle assembly is operated at room temperature. Whereas the rotational-line-intensity distribution of the spectrum recorded following expansion from the 300\,K nozzle is well described by a rotational temperature of about 150\,K, the spectrum of the $\mathrm{He_2^*}$ sample obtained with the 10\,K nozzle assembly is characterized by a bimodal rotational intensity distribution, with a cold component corresponding to a temperature of about 150\,K for levels with $N^{\prime\prime}\le 7$  and a much hotter (nonthermal) component corresponding to rotational levels in the range $N^{\prime\prime} = 9 - 21$ with a maximum at $N^{\prime\prime}= 17$. The 10\,K nozzle assembly thus leads to a rotationally warmer sample than the 300\,K nozzle assembly. This somewhat counter-intuitive observation may be explained by the large rotational constant ($B_0=7.10163(54)\,\mathrm{cm^{-1}}$, see Ref.~\cite{raunhardt08a}) of $\mathrm{He_2^*}$ and the fact that only odd-$N^{\prime\prime}$ rotational levels of the $\mathrm{a\,^3\Sigma_u^+}$ state are allowed by the generalized Pauli principle: The spacing between successive rotational levels rapidly becomes larger than the thermal collision energy when the nozzle assembly is held at low temperature and, consequently, cooling of the rotational degrees of freedom becomes very inefficient. The broad range of rotational states that are populated in the slow beam produced with the cold nozzle opens up the possibility to study levels with rotational energy exceeding 3000\,$\mathrm{cm^{-1}}$ by high-resolution spectroscopy under the collision-free conditions of a supersonic beam.

%%%%%%%%%%%%%%%%%%%%%%%%%%%%%%%%%%%%%%%%%%%%%%%%%%%%%%%%%%%%%%%%%%%%%%%%%%%
% DECELERATION
%%%%%%%%%%%%%%%%%%%%%%%%%%%%%%%%%%%%%%%%%%%%%%%%%%%%%%%%%%%%%%%%%%%%%%%%%%%

\subsection{Multistage Zeeman deceleration of $\mathrm{He}_2^*$}
\label{sec:results:decel}

\begin{figure}
\centering
\includegraphics{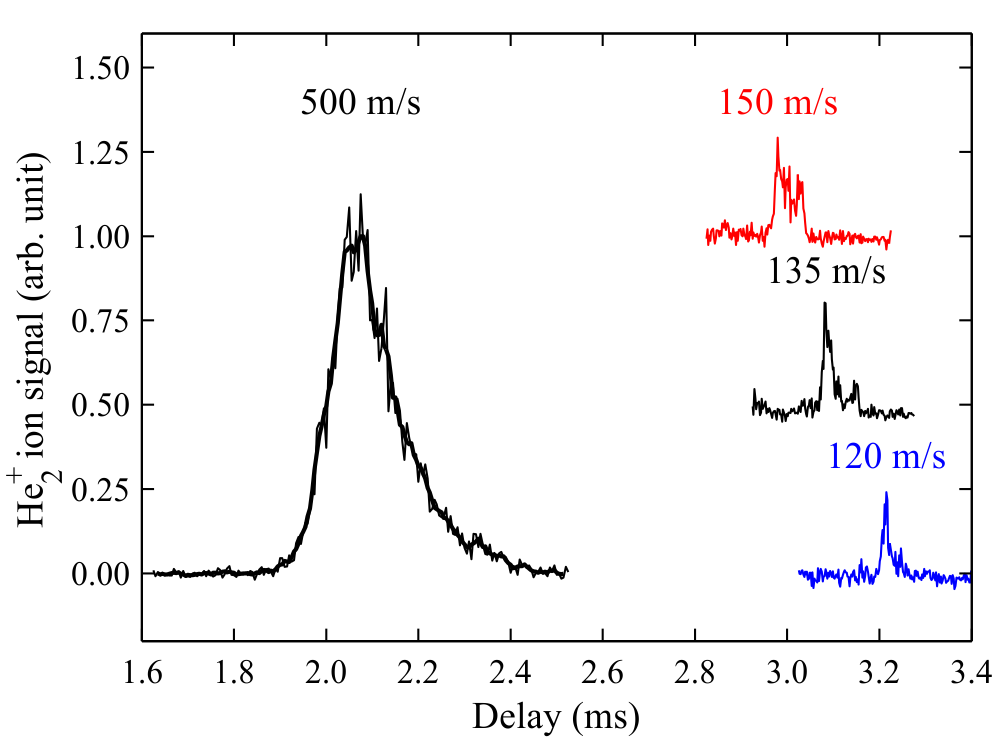}
\caption{Time-of-flight profiles of $\mathrm{He_2^*}$ detected 60\,mm beyond the last stage of the decelerator. The photoionization laser was set to 34467\,cm$^{-1}$, i.e., above all relevant ionization thresholds, to avoid the artificial enhancement of the photoionization signal of molecules in specific rotational levels of the $\mathrm{a\,^3\Sigma_u^+}$ state. The black trace with a broad TOF distribution centered at 500\,m/s corresponds to the undecelerated beam. The three distributions presented at times of flight between 2.9\,ms and 3.3\,ms were obtained using deceleration pulse sequences designed to decelerate molecules with initial velocities of 505\,m/s, 500\,m/s, and 495\,m/s, respectively, at a phase angle of 35$^\circ$.}
\label{fig:decel}
\end{figure}

When the nozzle assembly and the discharge electrodes are cooled down to about 10\,K, the He$_2^*$ molecules are entrained in a supersonic beam with a mean velocity of $\approx 500$\,m/s, which is well below the maximal initial velocity tolerated by our Zeeman decelerator. As mentioned in \refsec{sec:setup:zeeman}, 55 deceleration stages grouped in three modules suffice to decelerate He$_2^*$ to final velocities of 100\,m/s.

Time-of-flight profiles of $\mathrm{He_2^*}$ detected at the exit of the Zeeman decelerator, i.e., at a distance of about 1\,m from the nozzle orifice, are displayed in \reffig{fig:decel}. These profiles were obtained by the procedure (UV-laser photoionization mass spectrometry) used to characterize the beam source and discussed in \refsec{sec:results:source}. To prevent the artificial enhancement of the photoionization signal of molecules in specific rotational levels of the $\mathrm{a\,^3\Sigma_u^+}$ state, the UV-laser wave number was kept fixed at 34467\,cm$^{-1}$, in a structureless region of the photoionization spectrum. In \reffig{fig:decel}, the black trace, with a broad peak of $\mathrm{He_2^*}$ molecules centered at a delay of 2.05\,ms, represents the neutral TOF distribution recorded without switching any of the deceleration solenoids and corresponds to $\mathrm{He_2^*}$ molecules propagating at the original beam velocity of $\approx$500\,m/s through the entire decelerator. When running the decelerator with current pulses of 250\,A and a phase angle of 35$^\circ$, the TOF profiles observed in the region between 2.9\,ms and 3.3\,ms were obtained. The profiles with maxima at 2.95\,ms (red trace), 3.1\,ms (black trace) and 3.25\,ms (blue trace) were obtained under identical conditions except that the pulse sequences used for the deceleration were precalculated to decelerate metastable molecules with initial velocities of 505\,m/s, 500\,m/s, and 495\,m/s, respectively. Analysis of the time-of-flight profiles with the methods described in Ref.~\cite{wiederkehr11a} indicates that the final velocities are 150\,m/s, 135\,m/s, and 120\,m/s, respectively, in agreement with the values predicted by numerical particle-trajectory simulations. The structure of the peaks corresponding to the decelerated molecules is a consequence of the evolution of the phase-space distribution in the decelerator~\cite{wiederkehr10b}.

The neutral-TOF distributions presented in \reffig{fig:decel} demonstrate the possibility to tune the final velocity of the decelerated beam in the range between 100 and 150\,m/s. Although the present setup allows for the deceleration to lower final velocities in a straightforward way, either by operating the decelerator at a larger phase angle or by applying larger deceleration magnetic fields, the detection of these slow molecules becomes increasingly difficult. In particular, the rather large distance of 60\,mm between the last solenoid of the decelerator and the detection volume inevitably reduces the particle density by transverse defocussing and longitudinal dispersion effects, which play a larger role at lower final velocities~\cite{wiederkehr10b}. For the efficient generation of beams with velocities below 100\,m/s, deceleration pulse sequences with optimized phase angles, such as those already used for trap-loading schemes~\cite{wiederkehr10a}, might be useful to focus and bunch the molecules into the detection region.

%%%%%%%%%%%%%%%%%%%%%%%%%%%%%%%%%%%%%%%%%%%%%%%%%%%%%%%%%%%%%%%%%%%%%%%%%%%
% SPECTROSCOPY OF DECELERATED HE2
%%%%%%%%%%%%%%%%%%%%%%%%%%%%%%%%%%%%%%%%%%%%%%%%%%%%%%%%%%%%%%%%%%%%%%%%%%%

\subsection{Spectroscopy and rotational-state distribution of the decelerated $\mathrm{He_2^*}$ beam}
\label{sec:results:spec}

\begin{figure*}
\centering
\includegraphics{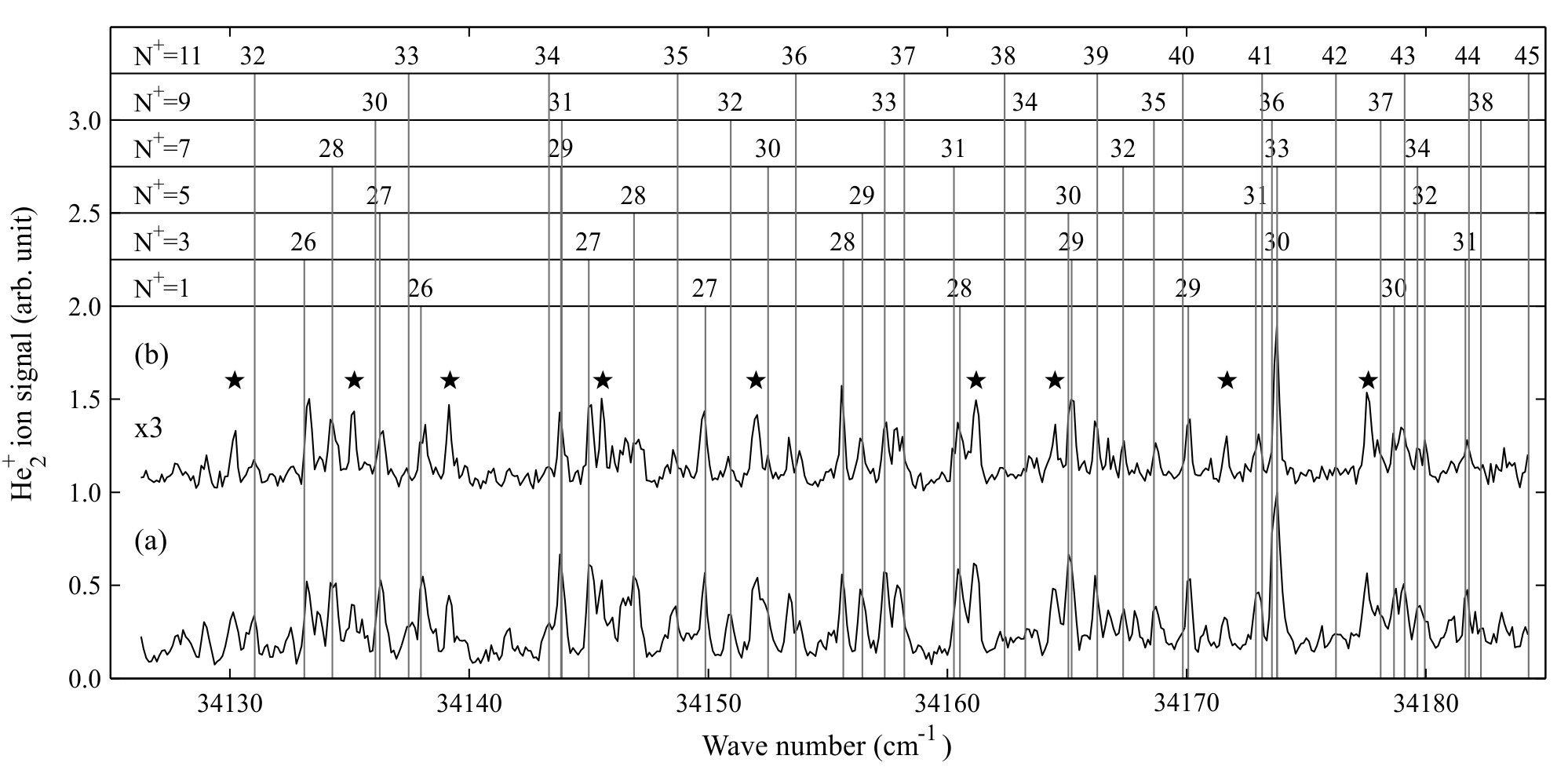}
\caption{Photoionization spectra of $\mathrm{He_2}$ in the vicinity of the ionization threshold from the $\mathrm{a\,^3\Sigma_u^+}$ state recorded for (a) an undecelerated beam and (b) a beam decelerated to a final velocity of 135\,m/s. The assignment bars indicate the positions of the $N^{\prime\prime}\rightarrow np(N^+=N^{\prime\prime})_{(N=N^+)}$ Rydberg states for $N^+=1-9$. The lines marked by asterisks belong to $N=N^+ \pm 1$ Rydberg series which are perturbed by rotational channel interactions. The intensity of trace (b) was multiplied by a factor of three and vertically offset for clarity.}
\label{fig:spectrum}
\end{figure*}

The maximum intensity of the peak in \reffig{fig:decel} corresponding to the decelerated He$_2^*$ molecules is only a factor of three weaker than the maximum intensity of the undecelerated beam. Consequently, the internal-state distribution of the decelerated sample can be characterized by spectroscopic methods with almost the same sensitivity as achieved for the undecelerated beam (see \refsec{sec:results:source}). Unfortunately, it was not possible to record PFI-ZEKE photoelectron spectra in the region beyond the last coil of the decelerator. Indeed, the detection of the photoelectrons following extraction in the direction parallel to the beam propagation axis was hindered by the background signal originating from metastable atoms and molecules and electrons generated by the discharge. The stray magnetic fields along the decelerator axis further prevented the detection of electrons extracted in the direction perpendicular to the beam propagation axis. The internal-state distribution of the decelerated sample had therefore to be determined from the photoionization spectrum of $\mathrm{He_2^*}$, which is dominated by contributions from autoionization resonances in the vicinity of the ionization thresholds (see Fig.~9 of reference~\cite{raunhardt08a}).

The photoionization spectra recorded in the region 34125--34185\,$\mathrm{cm}^{-1}$ for decelerated and undecelerated $\mathrm{He_2^*}$ beams are compared in \reffig{fig:spectrum}. These spectra are dominated by $N^{\prime\prime}\rightarrow n{\rm p} (N^{+}=N^{\prime\prime})_{(N=N^{+},N^+\pm 1)}$ Rydberg series~\cite{raunhardt08a, sprecher14a}. Whereas the $n{\rm p} N^{+}_{(N=N^+)}$ series, for which the character of the Rydberg electron is  $\pi_u^-$, are regular and well described by Rydberg's formula, the $n{\rm p}N^+_{(N=N^{+}-1)}$ and $n{\rm p}N^+_{(N=N^{+}+1)}$ series are of mixed $\sigma_u^+$ and $\pi_u^+$ character and are strongly perturbed by rotational channel interactions with the $n{\rm p}(N^+-2)_{(N=N^+-1)}$ and $n{\rm p}(N^++2)_{(N=N^++1)}$ series, respectively. For clarity, the assignments are only given for the regular series in \reffig{fig:spectrum} and the lines belonging to the perturbed series, which have also been assigned using the multi-channel-quantum-defect-theory models described in Refs.~\cite{raunhardt08a, sprecher13a}, are marked by asterisks.

The unambiguous assignment of transitions originating from rotational levels of metastable $\mathrm{He_2^*}$ with $N^{\prime\prime}$ values between 1 and 9 demonstrates that many rotational levels are populated in the decelerated sample. Moreover, the almost identical appearance of the spectra of the decelerated and undecelerated samples presented in \reffig{fig:spectrum} implies that the multistage Zeeman deceleration process is not rotationally state selective in He$_2^*$. This observation stands in stark contrast to what is observed in the multistage Zeeman deceleration of $\mathrm{O_2}$, for which the decelerated sample was found to consist exclusively of molecules in the $M_J=2$ magnetic sublevel of the $J=2$ fine-structure component of the $N=1$ rotational ground state~\cite{wiederkehr12a}. The reasons for this very different behavior lie in the very different relative magnitudes of the rotational constants and spin-rotation splittings of these two molecules, and in the fact that the Paschen-Back limit of the Zeeman effect is reached at much lower magnetic fields in the metastable $\mathrm{a\,^3\Sigma_u^+}$ state of $\mathrm{He_2}$ than in the $\mathrm{X\,^3\Sigma_g^-}$ ground state of $\mathrm{O_2}$, as is discussed in more detail in the next section.

%%%%%%%%%%%%%%%%%%%%%%%%%%%%%%%%%%%%%%%%%%%%%%%%%%%%%%%%%%%%%%%%%%%%%%%%%%%
% DISCUSSION AND CONCLUSION
%%%%%%%%%%%%%%%%%%%%%%%%%%%%%%%%%%%%%%%%%%%%%%%%%%%%%%%%%%%%%%%%%%%%%%%%%%%

\section{Discussion and conclusions}
\label{sec:conclusion}

\begin{figure}
\centering
\includegraphics{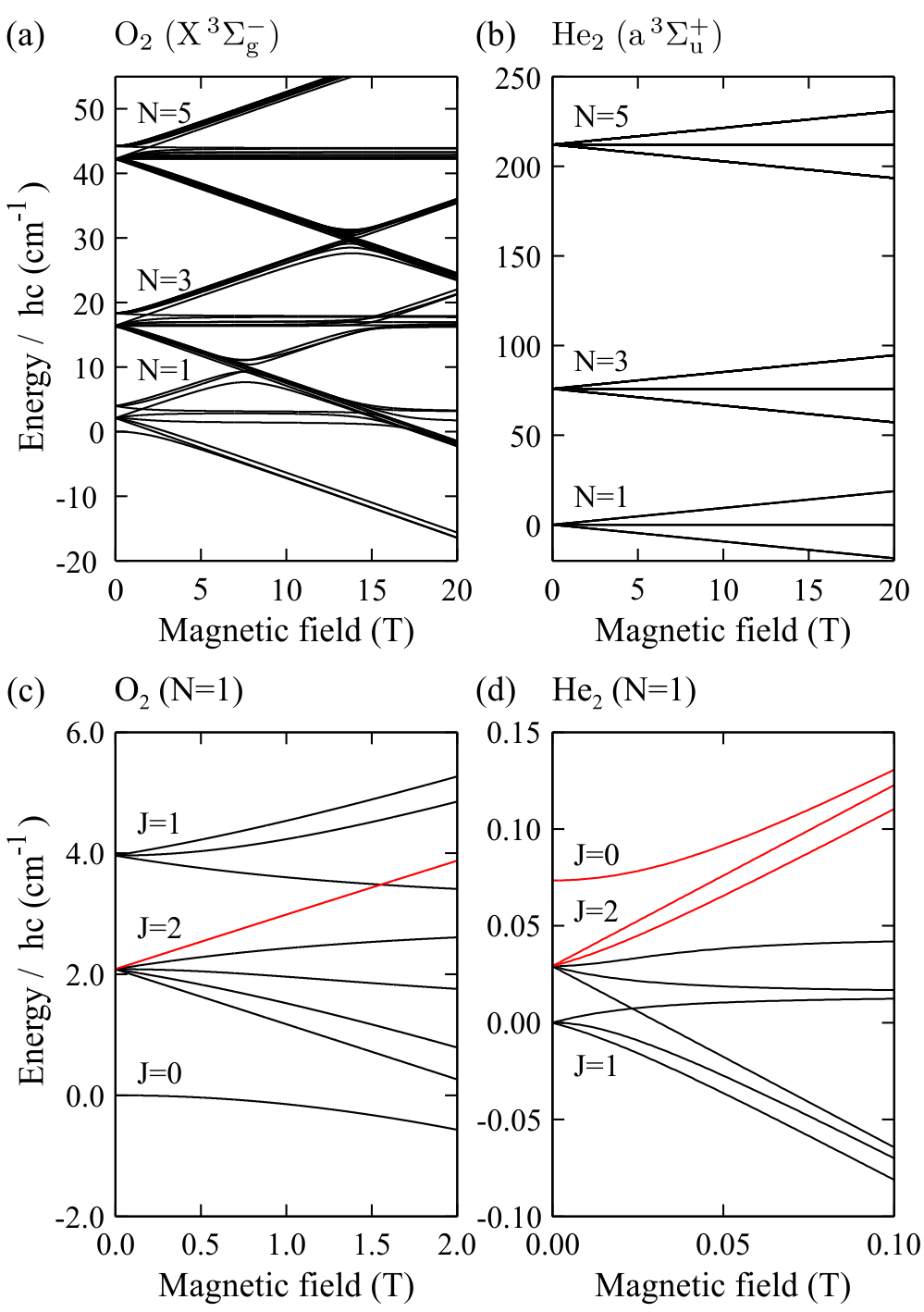}
\caption{Comparison of the Zeeman effect in the $\mathrm{X\,^3\Sigma_g^-}$ ground state of $\mathrm{O_2}$ (left column) and in the $\mathrm{a\,^3\Sigma_u^+}$ state of $\mathrm{He_2}$ (right column).
Top row: Zeeman energy of the rotational states $N^{\prime\prime}=1-5$ in the magnetic-field range between 0 and 20\,T.
Bottom row: Zeeman energy in the magnetic-field range where the Zeeman shifts are comparable to the spin-rotation splittings. Note the different scales used in these panels.
}
\label{fig:zeemancomparison}
\end{figure}

In this article, the design and operational characteristics of a source of slow, velocity-tunable beams of metastable $\mathrm{He_2}$ molecules ($\mathrm{He_2^*}$) have been presented. The molecules are formed in a discharge, entrained in a supersonic expansion of He, and decelerated to low velocity in a multistage Zeeman decelerator. At the end of the decelerator, the density of $\mathrm{He_2^*}$ is sufficiently high for measurements of the Rydberg spectrum of $\mathrm{He_2}$ by photoionization spectroscopy.

$\mathrm{He_2}$ in its metastable $\mathrm{a\,^3\Sigma_u^+}$ state has almost 20\,eV of internal energy, which is enough to ionize any atom or molecule except He. Consequently, it is not possible to form a supersonic beam of $\mathrm{He_2^*}$ seeded in any carrier gas other than atomic helium. A supersonic beam of pure He expanding from a room-temperature reservoir held at high stagnation pressure, however, has  a mean velocity of $\approx$2000\,m/s, which is beyond the range of velocities for which deceleration is possible with our multistage Zeeman decelerator~\cite{wiederkehr11a}. Indeed, such a beam propagates by more than the 7.2\,mm length of the solenoids used for deceleration during the switch-off time of about 8\,$\mu$s of the current pulses. To overcome this limitation, we have developed a source which can be cooled to 10\,K and at the exit of which $\mathrm{He_2^*}$ is generated in a discharge. $\mathrm{He_2^*}$ is then entrained by the He carrier gas in a supersonic beam having a mean velocity of about 500\,m/s. Phase-stable deceleration to velocities as low as 100\,m/s could be achieved using a Zeeman decelerator consisting of 55 stages. The velocity distribution of the decelerated molecules was measured by time-of-flight spectrometric methods and their internal-state distribution was determined from the analysis of well-resolved Rydberg series converging on the rovibrational levels of the $\mathrm{X\,^2\Sigma_u^+}$ ground electronic state of $\mathrm{He_2^+}$. The low translational temperature of the decelerated sample (less than 100\,mK) and the broad range of populated rotational levels ($N^{\prime\prime}$ ranging from 1 to 21) makes this source ideally suited to measurements of the Rydberg spectrum of $\mathrm{He_2}$, of the ionization energy of $\mathrm{He_2^*}$ and of the rovibrational structure of $\mathrm{He_2^+}$ by high-resolution photoionization and photoelectron spectroscopic methods.

Unexpectedly, the population of metastable molecules in the supersonic expansion were found to be distributed over many more rotational levels when the valve was cooled down to low temperature than when it was operated at room temperature. We interpret this observation as arising from the fact that the rotational spacings in $\mathrm{He_2^*}$ are much larger than the thermal collision energy when the nozzle assembly is held at low temperature, which effectively hinders rotational cooling. The broad range of rotational levels in the supersonic expansion enabled us to study how the Zeeman deceleration process affects the rotational-state distribution. In contrast to O$_2$, for which multistage Zeeman deceleration led to the formation of slow beams of molecules in a single magnetic sublevel of a single spin-rotational level of the ground electronic state~\cite{wiederkehr12a}, we did not observe any rotational state selectivity in the multistage Zeeman deceleration of $\mathrm{He_2^*}$. The reason for the very different behaviors observed in O$_2$ and $\mathrm{He_2^*}$ lies in the different nature of the Zeeman effect in these two molecules, which itself results from the fact that $\mathrm{He_2^*}$ has a much larger rotational constant than O$_2$ ($B_0=7.10163(54)\,\mathrm{cm^{-1}}$~\cite{raunhardt08a} vs. $1.43767638(55)\mathrm{cm^{-1}}$~\cite{huber79a, tomuta75a}) and spin-rotational splittings more than two orders of magnitude smaller than $\mathrm{O_2}$~\cite{focsa98a}.

The Zeeman effect in the ground vibrational level of $\mathrm{O_2}$ $\mathrm{X\,^3\Sigma_g^-}$ and $\mathrm{He_2}$ $\mathrm{a\,^3\Sigma_u^+}$ are compared in \reffig{fig:zeemancomparison}. In $\mathrm{O_2}$, the Paschen-Back regime is not reached at the magnetic fields up to 2\,T used in our experiments because of the large fine-structure splitting of the rotational levels. Moreover, interactions between magnetic sublevels of the same $M_J$ values but belonging to different rotational levels cause strong nonlinearities and avoided crossings already below 5\,T. In $\mathrm{He_2^*}$, the Paschen-Back regime is already reached at fields of less than 0.1\,T, and the large rotational spacings prevent any avoided crossings even at fields in excess of 15\,T. Consequently the low-field-seeking magnetic sublevels of the different rotational levels of $\mathrm{He_2^*}$ are subject to almost identical Zeeman shifts at the fields used in our deceleration experiments, which makes the deceleration process completely insensitive to the degree of rotational excitation.

Consideration of the Zeeman effect in the rotational levels of the $\mathrm{a\,^3\Sigma_u^+}$ of $\mathrm{He_2}$, however, reveals that none of the magnetic sublevels of the spin-rotational levels with $J^{\prime\prime}=N^{\prime\prime}$ are low-field seeking (see \reffig{fig:zeemancomparison}~(d)). Multistage Zeeman deceleration of $\mathrm{He_2^*}$ therefore acts as a filter which selects the two spin-rotational components with $J^{\prime\prime}=N^{\prime\prime}\pm 1$ and rejects the third. This property will represent an advantage in studies of the very congested Rydberg spectrum of $\mathrm{He_2}$.

%%%%%%%%%%%%%%%%%%%%%%%%%%%%%%%%%%%%%%%%%%%%%%%%%%%%%%%%%%%%%%%%%%%%%%%%%%%
% ACKNOWLEDGMENTS
%%%%%%%%%%%%%%%%%%%%%%%%%%%%%%%%%%%%%%%%%%%%%%%%%%%%%%%%%%%%%%%%%%%%%%%%%%%

\begin{acknowledgments}
We thank Richard Maceiczyk and Patrick Seewald for their support in the early phase of this work. This work is supported by the Swiss National Science Foundation (Project No.\ 200020-149216) and the European Research Council advanced grant program (Project No.\ 228286). M.~M.\ thanks ETH Z\"urich for the support through an ETH fellowship.
\end{acknowledgments}

%%%%%%%%%%%%%%%%%%%%%%%%%%%%%%%%%%%%%%%%%%%%%%%%%%%%%%%%%%%%%%%%%%%%%%%%%%%
% BIBLIOGRAPHY
%%%%%%%%%%%%%%%%%%%%%%%%%%%%%%%%%%%%%%%%%%%%%%%%%%%%%%%%%%%%%%%%%%

\end{document}